\documentclass[aps,pra,eqsecnum,twocolumn]{revtex4}

\usepackage{graphicx}

\usepackage{graphicx}

\newcommand{\be}{\begin{equation}}
\newcommand{\ee}{\end{equation}} 
\newcommand{\ba}{\begin{eqnarray}}
\newcommand{\ea}{\end{eqnarray}}

\def\opone{\leavevmode\hbox{\small1\kern-3.8pt\normalsize1}}
\def\l{\label}
\def\c{\cite}
\def\r{\ref}
\def\la{\langle}
\def\ra{\rangle}
\def\e{{\rm e}}
\def\i{{\rm i}}
\def\f{\frac}
\def\s{\sqrt}

\begin{document}

\title{Universality of Decoherence in the Macroworld}

\author{Walter T. Strunz}
\affiliation{Fakult\"at f\"ur Physik, 
Universit\"at Freiburg, Hermann-Herder-Str.3,
79104 Freiburg, Germany}
\author{Fritz Haake}
\affiliation{Fachbereich Physik, Universit\"at Essen,
45117 Essen, Germany}
\author{Daniel Braun}
\affiliation{Fachbereich Physik, Universit\"at Essen,
45117 Essen, Germany}

%
\date{\today}

\begin{abstract}

We consider environment induced decoherence of quantum superpositions to
mixtures in the limit in which that process is much faster than
any competing one generated by the Hamiltonian $H_{\rm sys}$
of the isolated system. 
This interaction dominated decoherence limit  
has previously not found much attention even though it is of importance
for the emergence of classical behavior in the macroworld, since 
it will always be the relevant regime for large enough
separations between the superposed wave packets.
The usual golden-rule 
treatment then does not apply but we can employ a short-time expansion for 
the free motion while keeping the interaction $H_{\rm int}$ in full. We 
thus reveal decoherence as a universal short-time phenomenon largely 
independent of the character of the system as well as the bath and of 
the basis the superimposed states are taken from.
Simple analytical expressions for
the decoherence time scales 
are obtained in the limit in which decoherence is even
faster than any timescale emerging from 
the reservoir Hamiltonian $H_{\rm res}$.

\end{abstract}

\maketitle

\section{Introduction}

Interferences from quantum superpositions of wave packets representing,
say, the translational motion of a body, become more and more difficult
to observe as the body becomes more massive and the superposed states
are made more distinct. Eventually, when the separation of wave packets
is increased towards macroscopic magnitudes (for which latter case we
shall speak of ``macroscopic superpositions''), classical behavior,
i.e. loss of the ability to interfere,
emerges. Somehow, along the way from microscopic to macroscopic
superpositions, the quantum capability of a particle to show up ``here''
and ``there'' simultaneously escapes detectability.

Two reasons are known for the elusiveness of macroscopic superpositions.
One of these even has a classical wave analogue. To explain it,
let us imagine a plane wave with (de Broglie or classical) wavelength
$\lambda$ traversing a spatial structure of linear dimension $d$
which splits the wave into partial ones. The parameter
$\lambda/d$ then determines the resolvability of interference
effects. For instance, in a double-slit experiment an
incoming plane wave gives rise to an outgoing interference pattern of
angular aperture $\lambda/d$. The latter angle becomes exceedingly
small when $\lambda$ is the de Broglie wavelength of a macroscopic body.

The second reason for the notorious absence of quantum superpositions
from the macroworld, called environment induced decoherence
\cite{Zeh,Zurek}, is of
dissipative origin and is the one of concern to us here. Decoherence is, for
microscopic bodies, just a facet of dissipation caused by
interactions with many-freedom surroundings. However, if two sufficiently
distinct wave packets $|\varphi_1\rangle,|\varphi_2\rangle$ are brought to
an initial superposition
$|\,\rangle=c_1|\varphi_1\rangle+c_2|\varphi_2\rangle$, the
density operator $\rho(t)$ starts out as the projector
$\rho(0)=|\,\rangle\langle\,|$ and then, for suitable coupling to the
environment (see below), decoheres to the mixture
$|c_1|^2|\varphi_1\rangle\langle\varphi_1|+|c_2|^2|\varphi_2\rangle\langle
\varphi_2|$, with the weights
$|c_i|^2$ still as in the initial superposition, on a time scale
$\tau_{\rm dec}$ while the subsequent relaxation of that mixture
 has a much longer characteristic
time $\tau_{\rm diss}$. The smallness of the decoherence time $\tau_{\rm
dec}$ is manifest in its proportionality to a power of Planck's constant
and inverse proportionality to a power of the ``distance'' $d$ between
the superposed packets,
\begin{equation}
  \tau_{\rm dec}\propto \frac{\hbar^\mu}{d^\nu}\quad {\rm with}\quad
  \mu,\nu>0\,.
  \label{1.1}
\end{equation}
We may interpret that power law as assigning a quantum scale of
reference $\propto \hbar^{\mu/\nu}$ to the distance $d$ such that
the decoherence time $\tau_{\rm dec}$ becomes vanishingly small when $d$
assumes classical magnitude. On the other hand, the characteristic times
for temporal changes of probabilities or other observables capable of a
well defined classical limit remain finite in the formal limit
$\hbar\to 0$. As a consequence, a given environment may have so weak an
influence that probability relaxation is hard to follow because of
$\tau_{\rm diss}$ being very large, while giving rise to unresolvably
small life times $\tau_{\rm dec}$ to coherences between sufficiently far
apart wave packets.

A variety of experimental studies of decoherence have been undertaken
\cite{Zeili,Welschi,Winey,Delft,Stony}, all of them involving weakly
coupled environments (``reservoirs'' or ``heat baths'') and wave
packet separations of but modest magnitudes: the acceleration of
decoherence over dissipation was not at all extreme, the time scale
ratio $\tau_{\rm dec}/\tau_{\rm diss}$ not even down to $10^{-2}$ yet.
Moreover, dissipation was sufficiently weak in all these experiments for
the decoherence time to exceed the time scales $\tau_{\rm
sys}$ characteristic of the free motion of the system isolated from the
environment. In that limit, ``a lot of'' free motion takes place
during decoherence, and therefore the latter process becomes rather system
specific in its characteristics. A unified treatment can, however, be
based on the very fact that the environmental influence is weak and thus
allows for perturbative treatment by the golden rule.

To illustrate decoherence in the golden-rule limit $\tau_{\rm
sys}<\tau_{\rm dec}< \tau_{\rm diss}$ one often considers a harmonic
oscillator of mass $M$ and frequency $\Omega$ and a bath in thermal
equilibrium. If the interaction
Hamiltonian is the product of two coupling agents, one for the system
($Q$) and the other for the bath ($B$), i.e. $H_{\rm int}=QB$, and
if two superposed wave packets are distinguished by the coupling agent
$Q$ in terms of the distance
$d=|q_1-q_2| 
= \Big|\la\varphi_1|Q|\varphi_1\ra-\la\varphi_2|Q|\varphi_2\ra\Big|$,
the golden rule is easily seen to yield the decoherence and
dissipation times
\ba
\f{1}{\tau_{\rm dec}^{\rm GR}}&=&\f{(q_1-q_2)^2}{\hbar^2}\int_0^\infty\!dt\,
\la\textstyle{\f{1}{2}}\{\tilde{B}(t),B\}\ra\, \cos{\Omega t}\,
\nonumber
\l{1.2} \\
\f{1}{\tau_{\rm diss}^{\rm GR}}&=&\frac{1}{M\Omega}\int_0^\infty\!dt\,
\la\textstyle{\f{\i}{\hbar}}[\tilde{B}(t),B]\ra \,\sin{\Omega t}\,,
\ea
where $\tilde{B}(t)=\e^{\i H_{res}t/\hbar}B\e^{-\i H_{res}t/\hbar}$
refers to free time evolution of the bath; note that the dissipation
time involves the response function
$\la\textstyle{\f{\i}{\hbar}}[\tilde{B}(t),B]\ra$ and the decoherence
time the equilibrium correlation function
$\la\textstyle{\f{1}{2}}\{\tilde{B}(t),B\}\ra$, with $\{\cdot,\cdot\}$
and $[\cdot,\cdot]$ denoting anticommutator and commutator,
respectively, and $\la\ldots\ra$ thermal equilibrium average.
Interestingly, the golden-rule decoherence time obeys the power law
(\ref{1.1}) while the dissipation time is independent of Planck's
constant and of the distance $d$.

Our principal goal in the present paper is to contrast the golden-rule limit
$\tau_{\rm sys}<\tau_{\rm dec}< \tau_{\rm diss}$ with the opposite case
in which decoherence is the fastest process by far,
\begin{equation}
  \tau_{\rm dec}\ll\tau_{\rm sys}, \tau_{\rm diss},
  \label{1.3}
\end{equation}
irrespective of the relative size of $\tau_{\rm sys}$ and $\tau_{\rm
diss}$. That {\it interaction dominated limit} prevails for sufficiently
far apart wave packets and, in particular, for the decoherence of truly
macroscopic superpositions; it may therefore be seen as relevant for the
emergence of classical behavior in the macroworld
and for the difficulties in experimentally 
pushing quantum coherent dynamics
into the macroscopic domain.
Moreover, the limit (\ref{1.3}) must
assign much more universal properties to decoherence since it allows no
or ``very little'' free motion during times of the order $\tau_{\rm
dec}$. We shall, in fact, see that our interaction dominated limit
(\r{1.3}) yields decoherence times independent of the force $F(Q)$ that
may act on the isolated body. The decoherence times to be met with will
involve different exponents $\mu,\nu$ in (\r{1.1}) than the
golden-rule one, $\tau_{\rm dec}^{\rm GR}$ of (\r{1.2}).

For a major part of the paper we not only base our
analysis on (\r{1.3}), but 
furthermore assume
\begin{equation}
  \tau_{\rm dec}\ll\tau_{\rm res},
  \label{1.4}
\end{equation}
i.e. decoherence is fast even on environmental time scales. 
In that case, simple expressions of
universal character, independent
of the details of environmental dynamics, are obtained.

As soon as we drop (\r{1.4}) yet retain (\r{1.3}), we find more
complicated decoherence dynamics, the temporal decay
now being governed by the details of the time evolution of
environmental correlations.

It would be highly desirable to experimentally observe the crossover
from the golden-rule limit to our interaction dominated limit (\r{1.3}),
and further to the extreme limit where both (\r{1.3}) and (\r{1.4})
are satisfied.
As already mentioned above, the experiments done thus far pertain to the
golden-rule limit where the separation exponent $\nu$
takes on the value 2. We
shall present some discussion of the crossover condition in 
the accompanying paper \cite{oscillator}.
A quantitative treatment of that crossover itself will have
(i) to be non-perturbative (like ours and in contrast to the golden
rule) and (ii) have to avoid even the short-time approximation w.r.t.
free motion whose simplicity we will take profit of in the
present paper. In the accompanying paper \cite{oscillator}
we treat the crossover in question for
an exactly solvable model where both the system and the bath consist
of harmonic oscillators.

The above remark about interaction dominated decoherence showing greater
universality than its golden-rule counterpart deserves some
qualification. If both
(\r{1.3}) and (\r{1.4}) are
satisfied, of the three parts of the Hamiltonian of the composite
system, $H=H_{\rm sys}+H_{\rm res}+H_{\rm int}$, the generators of free
motion, $H_{\rm sys}$ and $H_{\rm res}$, play but a minor role in
comparison to the interaction part $H_{\rm int}$. As a consequence, it
is not of much importance whether the ``body'' under study is an
oscillator, a large angular momentum or some other few-freedoms system.
All we require is the possibility of superposing wave packets with large
separations $d$, large in relation to microscopic quantum
scales. Similarly, it does not matter whether the reservoir is composed
of harmonic oscillators (like modes of electromagnetic or elastic
waves), atoms or other entities; what counts is that the reservoir has
many degrees of freedom effective in $H_{\rm int}$, i.e. for $H_{\rm
int}=QB$ in its coupling agent $B$; we shall assume $B=\sum_{i=1}^N B_i$
with $N$, the number of reservoir freedoms, large.

It is appropriate to admit that in one other respect our interaction
dominated limit is no more but rather even a bit less universal than the
golden-rule one. Obviously, the system coupling agent $Q$ in $H_{\rm
int}=QB$ is distinguished over other system observables not showing up
in the interaction. As we shall see the coupling agent $Q$ is most
effective in decohering superpositions of wave packets with large
separations
$\Big|\la\varphi_1|Q|\varphi_1\ra-\la\varphi_2|Q|\varphi_2\ra\Big|$ and
considerably less effective if the distinction of the packets is one
w.r.t. some other observable $P$ not commuting with the coupling agent
$Q$. In fact, packets far apart in $Q$ will turn out to decohere with a
Gaussian decay of suitable indicators, like $\exp\{-(t/\tau_{\rm
dec}^Q)^2\}$, and that decay is captured already in zeroth order in
$H_{\rm sys}+H_{\rm res}$; in that zeroth order, however, wave packets
distinguished by $P$ but not by $Q$ would appear as retaining their relative 
coherence.
Such latter packets are in fact also decohered by $H_{\rm int}=QB$, but
in general in a non-Gaussian manner, like $\exp\{-(t/\tau_{\rm
dec})^n\}$ with $n>2$ and a decoherence time $\tau_{\rm dec}$ which
differs from $\tau_{\rm dec}^Q$ in the exponents $\mu,\nu$ but
still is of quantum character due to $\mu>0$; to capture that latter
decoherence, a ``little bit'' of free motion must be accounted for in a
systematic manner, as will be explained in Sect. \ref{IDD} below. On 
the other hand, decoherence fully symmetric in $Q$ and $P$ would 
result from an interaction involving both of these observables as 
coupling agents towards different reservoirs, $H_{\rm int}=QB_1+PB_2$, 
as described in Section V and previously pointed out in a first report on 
this project \cite{BHS}. 

The decay in terms of exponentials of powers $t^n$ just mentioned
arises when decoherence outruns both dissipation and the decay
of bath correlations, i.e., when both (\r{1.3}) and (\r{1.4}) 
are satisfied.
When only dissipation is cut short but bath correlation decay remains
effective, such that  (\r{1.3}) is respected but not (\r{1.4}),
the qualitative picture of the above discussion does not
change, yet the precise temporal course of decoherence involves the full time
dependence of the bath correlation function and can no longer be written
as an exponential of a power of time \cite{BHS}; Section VI of the present 
paper is devoted to that case. 

In Section VII we treat decoherence 
of superpositions of angular-momentum coherent states, with one component, 
$J_x$, of an angular momentum $\vec{J}$ acting as coupling agent. In analogy 
to our findings for superpositions of states distinguished by $P$ and $Q$, we
shall be led to most rapid decoherence for pairs of states with differing 
mean values of the coupling agent and slowest decoherence for pairs of states 
with coinciding mean values for all system observables coupled to the coupling 
agent by the evolution generated by $H_{\rm sys}$; ``rapid'' and ``slow'' will 
again be quantified by the exponents $\mu,\nu$ in the power law (\r{1.1}).

Before going in medias res, some words about related literature are in order.
As already mentioned, the interaction dominated short-time limit of decoherence
has received little attention, in spite of its obvious relevance for the 
transition from quantum to classical behavior. The only exceptions we are 
aware of are Joos and Zeh's short time expansion of an entanglement
measure based on a Schmidt decomposition \cite{JoosZeh}
and various articles that
refer to the exactly solvable model of a harmonic oscillator 
coupled to a reservoir itself consisting of harmonic oscillators 
\c{Ullersma,HR,etc}. In a discussion of the quantum measurement process where
an entangled state of a microscopic quantum system and a macroscopic pointer 
involves superpositions of macroscopically distinct pointer states, Haake and 
{\.Z}ukowski \c{HZ} have employed the oscillator model and its exact solution 
to argue that the superposition in question decoheres in the limit  
$\tau_{\rm dec}\ll\tau_{\rm sys}$. More recently, the importance of that limit
has also been realized by Privman \c{Privman}. In a series of papers Ford, 
Lewis and O'Connell \c{FLOP} argue that two wave packets of widths $\sigma$ and 
separation $d$ in $Q$-space experience the time scale 
$\tau_{\rm FLO}=\sigma^2/dv$, where $v=\sqrt{k_BT/m}$ is a thermal velocity 
with $m$ a typical mass; they point out that the latter time may be short 
compared to the golden-rule prediction for the decoherence time. Interestingly,
$\tau_{\rm FLO}$ is independent of Planck's constant as well as of the strength
of any interaction $H_{\rm int}$. As the appearance of the 
temperature $T$ suggests and close inspection reveals, the underlying thermal 
ensemble embodies no coherence extending over the distance $d$ between 
the wave packets to begin with; instead, initial coherences are confined to 
the  thermal de Broglie wavelength characterizing the thermal ensemble. The
relaxation processes taking place on the time scale $\tau_{\rm FLO}$ are 
therefore not related to the decoherence of a macroscopic superposition to a 
mixture.

\section{Superpositions of Distinct Wave Packets}\label{packets}
 
We consider a single-freedom system for which the coordinate $Q$
and momentum $P$ obey the canonical commutation rule
\begin{equation}
[P,Q]=\frac{\hbar}{{\rm i}}\,. \label{2.1}
\end{equation}
The initial states we shall have to deal with are pure states
of the form of superpositions of two separate wave packets,
\be
|\ra=c_1|\varphi_1\ra+c_2|\varphi_2\ra\,,\quad |c_1|^2+|c_2|^2=1.
\label{2.2}
\ee
We may specify the individual packets in either the position or momentum 
representation and choose, for the sake of convenience, the Gaussians
\ba
\la q|\varphi_i\ra&=&\varphi_i(q)=\f{1}{(2\pi\sigma)^{1/4}}\,\e^{\i
p_i(q-q_i)/\hbar}
\,\e^{-(q-q_i)^2/4\sigma}\nonumber\\
\la p|\varphi_i\ra&=&\tilde{\varphi}_i(p)=\f{(2\pi\sigma)^{1/4}}
{(\pi\hbar)^{(1/2)}}\,\e^{-\i pq_i/\hbar}
\,\e^{-\sigma(p-p_i)^2/\hbar^2}\,,
\label{2.3}
\ea
with \quad i=1,2\,.Needless to say, $\varphi_i(q)$ and
$\tilde{\varphi}_i(p)$ are Fourier transforms of one another.
These packets are located in position space at $q_i$ with (rms)
uncertainty $\Delta q=\s{\sigma}$ and in momentum space at $p_i$ with
uncertainty $\Delta p=\hbar/2\s{\sigma}$; the uncertainty product
$\Delta q\Delta p=\hbar/2$ is the minimum one allowed by the uncertainty
principle; were we to choose $\sigma$ as a classical quantity
independent of Planck's constant, we would confront two extremely
squeezed states with the momentum much more sharply defined than the
position; we will actually envisage the symmetric situation
$\sigma\propto\hbar$ where both $\Delta q$ and $\Delta p$ are
$\propto\s{\hbar}$, like for coherent states \cite{Glauber}. 
To ensure good separation
we stipulate that either $\Delta q\ll|q_1-q_2|$ or $\Delta
p\ll|p_1-p_2|$ or both (see Fig. 1). Actually, inasmuch as we are
interested in ``macroscopic superpositions'' we may assume at least one
of the two distances $|q_1-q_2|,|p_1-p_2|$ of classical magnitude,
i.e. independent of $\hbar$.

\begin{figure}\label{fig0}
\includegraphics[angle=270,scale=.25]{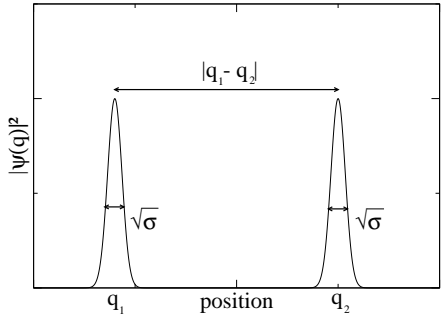}
\caption{Position space density $|\psi(q)|^2$ for a coherent
superposition of two Gaussian wave packets as envisaged in
this paper. The distance between the packets is assumed much
larger than their individual spread.}
\end{figure}

Our choice of Gaussian packets is a matter of convenience; it will allow
us to evaluate all subsequently encountered integrals analytically. The
universal decoherence laws to be established rest on sufficient
separation of the two packets, however, rather than on their specific
form or their minimum-uncertainty property.

The initial density operator corresponding to the state (\ref{2.2}) is a sum 
of four terms,
\be
\rho_{\rm sys}(0)=\sum_{i,j=1}^2c_ic_j^*\,|\varphi_i\rangle\langle\varphi_j|
=\sum_{i,j}c_ic_j^*\,\rho^{ij}_{\rm sys}(0)\,,
\l{2.4}
\ee
two ``diagonal'' ones weighted by probabilities $|c_i|^2$ and two
off-diagonal ``interference terms'' $\rho^{12}_{\rm
sys}(0)=|\varphi_1\ra\la\varphi_2|=\rho^{21}_{\rm sys}(0)^\dagger$
weighted by the ``coherences'' $c_1c_2^*$ and $c_1^*c_2$.

Inasmuch as quantum mechanical time evolution is represented by linear
operators each of the four terms in (\ref{2.4}) has its own temporal
successor $c_ic_j^*\rho^{ij}_{\rm sys}(t)$. To show that interaction
with an environment tends to destroy the interference terms before the
diagonal terms change noticeably we shall employ the norms
\be
N_{ij}(t)={\rm Tr_{sys}}\, \rho^{ij}_{\rm sys}(t)\rho^{ij}_{\rm sys}(t)^\dagger
\l{2.5}\,.
\ee
Clearly, if the system in question were isolated these norms would all
remain time independent, $N_{ij}(t)=1$, since the unitary time evolution
operators $U_{\rm sys}(t)=\e^{-\i H_{\rm sys}t/\hbar}$ would cancel
under the trace operation. The time scale separation we are after arises
only due to the interaction with an environment, and then only if the
initial wave packets $\varphi_i$ are sufficiently distinct.

\section{Interaction Dominated Decoherence}\l{IDD}

To allow for dissipative motion of $Q$ and $P$ we introduce
a reservoir with many degrees of freedom and deal with a Hamiltonian of
the structure $H=H_{\rm sys}+H_{\rm res}+H_{\rm int}$. We need not
specify the Hamiltonian $H_{\rm res}$ governing the
free motion of the environment; the Hamiltonian of the isolated
single-freedom system is taken as the usual sum of a kinetic and
a potential term, $H_{\rm sys}=P^2/2M+V(Q)$; for the interaction
Hamiltonian, however, we do assume a slightly restrictive form involving
only one of the two system observables, say $Q$, as a coupling
agent,
\begin{equation} H_{\rm int}=QB\,, \label{3.1}
\end{equation}
with $B$ some reservoir coupling agent which should
involve all degrees of freedom of the reservoir in a way to be commented
on below.

The simplest initial state to deal with has our single-freedom system prepared
so as to be statistically independent from the reservoir; the initial joint 
density operator then takes the form of a product,
\be
\rho(0)=\rho_{{\rm sys}}(0)\rho_{{\rm res}}(0)\,,
\l{3.2}
\ee
with $\rho_{\rm sys}$ representing the superposition of two distinct wave 
packets as described in the preceeding section. The initial density operator 
of the reservoir could but need not be the thermal equilibrium state with 
respect to $H_{\rm res}$; our precise demand on $\rho_{{\rm res}}(0)$ will be 
given presently.

The reduced system density operator originating from any one of the four terms
$\rho_{\rm sys}^{ij}(0)$ can now be written as
\be
\rho_{\rm sys}^{ij}(t)={\rm Tr_{res}}\e^{-\i Ht/\hbar}
\rho_{{\rm sys}}^{ij}(0)\rho_{{\rm res}}(0)\e^{\i Ht/\hbar}\,.
\l{3.3}
\ee 
In view of our intention to evaluate the norms $N_{ij}(t)$ defined in (\r{2.5}) it is
advantageous to pass to the interaction picture and write the time evolution operator as
\ba
\e^{-\i Ht/\hbar}&=&\e^{-\i (H_{\rm sys}+H_{\rm res})t/\hbar}\tilde{U}(t)=
U_0(t)\tilde{U}(t)\,,
\nonumber\\
\tilde{U}(t)&=&\Big(\!\e^{-\i\int_0^t\!dt'\tilde{H}_{\rm
int}(t')/\hbar}\Big)_{\!+}\,,
\l{3.4}\\
\tilde{H}_{\rm int}(t)&=&U_0^{\dagger}(t)H_{\rm int}\,U_0(t)
=\tilde{Q}(t)\tilde{B}(t) \,,\nonumber
\ea
where $(\ldots)_+$ demands time ordering of the operator product $(\ldots)$.
The norms in question thus read
\ba
N_{ij}(t)&=&{\rm Tr_{sys}}\, \tilde{\rho}_{\rm
sys}^{ij}(t)\tilde{\rho}_{\rm sys}^{ji}(t)\l{3.5}\\
\tilde{\rho}_{\rm sys}^{ij}(t)&=&{\rm Tr_{res}}\tilde{U}(t)\rho^{ij}_{\rm sys}(0)
\rho_{\rm res}(0)\tilde{U}^{\dagger}(t)\nonumber\,.
\ea

We are interested in the limiting case where decoherence, i.e. the decay
of $N_{12}(t)$ is faster than any process arising in the absence of the
coupling $H_{\rm int}$. 
Our results will selfconsistently confirm this limit as
relevant for large enough
separations (w.r.t position, or momentum, or both) 
between the
wave packets. The short-time behavior thus aimed
at allows to approximate the interaction-picture evolution operator
$\tilde{U}(t)$ by expanding its logarithm as a power series in the time
$t$. To find that expansion we start with the interaction picture
Hamiltonian
\begin{eqnarray}
\tilde{Q}(t)\tilde{B}(t)&=&
\left(Q+M^{-1}Pt-V'(Q)t^2/2+\ldots\right)\nonumber\\
& &\times
\left(B+\dot{B}t+\ddot{B}t^2/2+\ldots\right)
\,,
  \label{3.6}
\end{eqnarray}
where $\dot{B}=\frac{\i}{\hbar}[H_{\rm
res},B]\,,\ddot{B}=\frac{\i}{\hbar}[H_{\rm res},\dot{B}]$. 
Notice that this short-time expansion is meaningful only if both
conditions (\r{1.3}) and (\r{1.4}) are satisfied. We shall drop the latter
condition in Sect. \ref{shortmemory} where we keep the full time dependence
of $\tilde{B}(t)$.

A simple sequence of
unitary transformations, described in the appendix, brings
about the desired expansion as
\begin{equation}
  \tilde{U}(t)=\e^{-\frac{\i}{\hbar}\{Q(Bt+\dot{B}t^2/2)+PBt^2/2M+\ldots\}}
  \,,
  \label{3.7}
\end{equation}
where the dots refer to cubic and higher-order terms in $t$; in
particular, the force $-V'(Q)$ enters $\ln\tilde{U}(t)$ only in order
$t^3$.

We intend to evaluate the trace ${\rm Tr_{sys}}$ in the
$Q$-representation, where $Q|q\ra\!=q|q\ra$ and, with an arbitrary state
vector $|\psi\ra\,,$
$\,\la\psi|P|q\ra=\i\hbar\frac{\partial}{\partial
q}\la\psi|q\ra,\;\la q|P|\psi\ra=-\i\hbar\frac{\partial}{\partial q}\la
q|\psi\ra$. We thus have
\begin{eqnarray}
  \la q|\rho_{\rm sys}^{ij}(t)|q'\ra = D_{Q}(t)\;
\la q|\rho_{\rm sys}^{ij}(0)|q'\ra
\end{eqnarray}
with the decoherence factor
\ba
  D_{Q}(t)&=&{\rm Tr_{res}}\e^{\,-\frac{\i}{\hbar}(q-q')(Bt+\dot{B}t^2/2)
  -(\frac{\partial}{\partial q}+\frac{\partial}{\partial
  q'})Bt^2/2M} \,
  \rho_{\rm res}(0)\nonumber\\
  &=&\left\la
  \e^{\,-\frac{\i}{\hbar}(q-q')(Bt+\dot{B}t^2/2)
  -(\frac{\partial}{\partial q}+\frac{\partial}{\partial
  q'})Bt^2/2M}\right\ra \,;
\label{3.8}
\ea
the large angular brackets in the last member of the foregoing equation
denote an average w.r.t. the initial state of the reservoir.

At this point we need to specify the previously announced structure of the
reservoir coupling agent as {\it additively} comprising a large number
$N$ of degrees of freedom,
\begin{equation}
  B=\sum_{i=1}^NB_i\,.
  \label{3.9}
\end{equation}
Moreover, we require the reservoir initial state $\rho_{\rm res}$ to
involve those many degrees of freedom with sufficiently weak
correlations for the central limit theorem to hold for the statistical
behavior of $B$ as well as its time derivative $\dot{B}$. To avoid
unnecessarily voluminous expressions in the sequel we also stipulate
vanishing initial means of these observables, $\la B\ra=\la
\dot{B}\ra=0$. The exponent ${\cal B}\equiv
-\frac{\i}{\hbar}(q-q')(Bt+\dot{B}t^2/2)-
(\frac{\partial}{\partial q}+\frac{\partial}{\partial q'})Bt^2/2M$
in the reservoir expectation value in (\ref{3.8}) is
thus assigned Gaussian statistics according to
$\la\e^{\cal B}\ra=\e^{\frac{1}{2}\la {\cal B}^2\ra}$.
The decoherence factor (\ref{3.8}) then takes the form
\begin{eqnarray}
  D_Q(t)&=&\e^{-(q-q')^2\la(Bt+\dot{B}t^2/2)^2\ra/2\hbar^2}
\nonumber\\  
 & &\times\,\e^{\i(q-q')(\frac{\partial}{\partial q}
+\frac{\partial}{\partial q'})
  \la (Bt+\dot{B}t^2/2)Bt^2\ra/2M\hbar}\nonumber\\
  & &\times\,
  \e^{(\frac{\partial}{\partial q}+\frac{\partial}{\partial q'})^2
  \la B^2\ra t^4/8M^2}\,;
  \l{3.10}
\end{eqnarray}
it may be worth noting that we could write three separate exponentials
since the relative displacement and the center-of-mass momentum commute,
$[q-q',\frac{\partial}{\partial q}+\frac{\partial}{\partial q'}]=0$.
We shall save a lot of space and gain better transparency by retaining,
in each of the three exponentials in $D_Q(t)$,
only the respective leading-order terms in the time $t$; it will become clear
further below that nothing of relevance for the final result is thus
lost. A similar calculation in the $P$-basis yields
\begin{eqnarray}
  \la p|\rho_{\rm sys}^{ij}(t)|p'\ra&=&D_P(t) 
\la p|\rho_{\rm sys}^{ij}(0)|p'\ra\,,
  \label{3.11}\\
  D_P(t)&=&
  \e^{(\frac{\partial}{\partial p}+\frac{\partial}{\partial
  p'})^2\la B^2\ra t^2/2}
  \nonumber\\& &\times\,
  \e^{\i(p-p')(\frac{\partial}{\partial p}+\frac{\partial}{\partial p'})
  \la B^2\ra t^3/2M\hbar}
  \nonumber\\ & &\times\,
  \e^{-(p-p')^2 \la B^2\ra t^4/8M^2\hbar^2}\,;\nonumber
\end{eqnarray}
note that now we have retained only the ${\cal O}(t^2)$
term in the first exponential and the ${\cal O}(t^3)$ term in the second
exponential.

The asymmetry between $Q$ and $P$ in the matrix elements
(\ref{3.10}) and (\ref{3.11}) arises from the distinction of the
coordinate $Q$ as the system coupling agent in the interaction (\ref{3.1}).
By their asymmetric appearance these matrix elements already suggest
different temporal courses of decoherence for superpositions of wave
packets macroscopically distinguished in $Q$ and in $P$; that difference
will become yet easier to discern once we have evaluated the norms
(\ref{3.5}). Upon there inserting the matrix element (\ref{3.10}),
integrating by parts, and changing integration variables to relative and
center-of mass coordinate as $k=q-q',\overline{q}=\frac{1}{2}(q+q')$,
and $\frac{\partial}{\partial\overline{q}}=\frac{\partial}{\partial
q}+\frac{\partial}{\partial q'}\equiv \partial$
we get
\begin{eqnarray}
  N_{ij}&=&
  \int d\overline{q}\,dk
  \varphi_i^*(\overline{q}+k/2)\varphi_j(\overline{q}-k/2)D_Q^2(t)
  \nonumber\\& & \qquad\times\,
  \varphi_i(\overline{q}+k/2)\varphi_j^*(\overline{q}-k/2)\,,
  \label{3.12} \\
  D_Q^2(t)&=&\e^{-k^2\la B^2\ra t^2/\hbar^2}\;
  \e^{\i k\partial\la B^2\ra t^3/M\hbar} \;
  \e^{\partial^2\la B^2\ra t^4/4M^2 } \nonumber
  \,.
\end{eqnarray}
The second and third exponentials in the foregoing quantity $D_Q^2(t)$
are integral operators acting on the subsequent functions of the
center-of-mass variable $\overline{q}$ as, respectively,
$\e^{\Delta\partial}f(\overline{q})=f(\overline{q}+\Delta)$ and
$\e^{\tau\partial^2}f(\overline{q})=\int\!dx
(4\pi\tau)^{-1/2}\e^{(\overline{q}-x)^2/4\tau}f(x)$, i.e., like shift and
diffusion. Of course, apart from the change of variables just
indicated $D_Q^2(t)$ is nothing but the square of $D_Q(t)$ given in
(\ref{3.10}). After inserting the initial states (\ref{2.3}) and
doing the three Gaussian integrals over $\overline{q},k,x$ we
finally obtain the ``coherence norm'' $N_{12}(t)$ in its dependence on the 
time $t$
and the separations $q_1-q_2$  and $p_1-p_2$ of the two wave packets in
$Q$-space and $P$-space,
\ba
N_{12}(t)=& &\left\{1+4\sigma \la B^2\ra t^2/\hbar^2
+{\cal O}(t^4)\right\}^{-1/2}\nonumber\\
& &\times\exp\left\{-(q_1-q_2)^2\la B^2\ra t^2
/\hbar^2\right\}\l{3.13}\\
& &\times\exp\left\{-(q_1-q_2)(p_1-p_2)\la B^2\ra t^3/M\hbar^2\right\}
\nonumber\\
& &\times\exp\left\{-(p_1-p_2)^2\la B^2\ra t^4/4M^2\hbar^2\right\}\nonumber\\
& &\!\!\!\!\!\!\!\equiv{\cal P}(t)\,{\cal E}^Q(t)\,{\cal
E}^{QP}(t)\,{\cal
E}^P(t) \,; \nonumber \ea 
for typographical reasons we have not indicated the corrections 
$\propto t^{n+1}$
to the leading-order terms $t^n$ in the three exponentials; they 
are independent of the separations $q_1-q_2$
and $p_1-p_2$; neither do these separations enter the order-$t^4$
correction in the prefactor ${\cal P}(t)$.

We have thus established one of the central results of the present 
paper and proceed to a critical appreciation.

\section{Discussion of Interaction Dominated Decoherence}\l{dissIDD}

\subsection{Decoherence Time Scales}

If the two wave packets in our superposition differ both in
their center positions and momenta, the three exponentials in the
coherence norm $N_{12}(t)$ have the decay times
\begin{eqnarray}
 \tau_{\rm dec}^Q &=& \f{\hbar}{|q_1-q_2|\sqrt{\la B^2\ra}}\,,
 \nonumber\\
 \tau_{\rm dec}^{QP} &=& \left(\f{M\hbar^2}{|(q_1-q_2)(p_1-p_2)|\la
 B^2\ra}\right)^{\frac{1}{3}}\,,
 \label{4.1}\\
 \tau_{\rm dec}^P &=& \left(\f{4M^2\hbar^2}{(p_1-p_2)^2\la
 B^2\ra}\right)^{\frac{1}{4}}\,.
 \nonumber
\end{eqnarray}

All three of these decoherence times are quantum in character and tend
to vanish in the formal classical limit $\hbar\to 0$. They may be
considered ordered in their magnitudes by the respective powers of
Planck's constant \cite{Fuss}
$\tau_{\rm dec}^Q\propto \hbar,\tau_{\rm
dec}^{QP}\propto \hbar^{2/3},\tau_{\rm dec}^P\propto\hbar^{1/2}$. The
``distances'' $|q_1-q_2|$ and $|p_1-p_2|$ between the two wave packets
appear as referred to quantum scales and in those units tend to take on
huge values if mesoscopic or even macroscopic. At any rate, it is the
smallness of the decoherence times for which macroscopic superpositions
would have little chance to be detectable even if preparable.

Which of the three exponentials wins out in governing the decoherence of
macroscopically distinct packets depends on the distances $|q_1-q_2|$
and $|p_1-p_2|$; obviously, different cases arise, and these will be
dealt with individually below.

It may be worth noting that the powers of Planck's constant as well as
those
of the distances differ from the ones more familiar from the Golden-Rule
result (\r{1.2}).

\subsection{Universality and Limits of Validity}

Inasmuch as our result for the decay of coherence between the superposed
wave packets is based on a short-time expansion of the (logarithm of)
the time evolution operator $\tilde{U}(t)$, (cf (\r{3.4}),(\r{3.7})), we
have to emphasize its limit of validity. To appreciate that limit we
must realize that it is the free motion of the single-freedom system
and the reservoir which were treated as nearly ineffective during the
decoherence, while the interaction $H_{\rm int}$ was kept in full; in
particular, the first exponential, ${\cal E}^Q(t)$, in the coherence norm 
(\r{3.13}) can
immediately be checked to arise from entirely neglecting $H_{\rm
sys}+H_{\rm res}$ in (\r{3.4}) and thus taking $\tilde{U}(t)=\e^{-\i
H_{\rm int}t/\hbar}=\e^{-\i QBt/\hbar}$. Shouldering the burden of the
${\cal O}(t^2)$ terms in (\r{3.7}), which bring in
$\dot{Q}=\f{\i}{\hbar}[H_{\rm sys},Q]=P/M,\;\dot{B}=\f{\i}{\hbar}[H_{\rm
res},B]$ is necessary only in the case $q_1=q_2$; note again that the
potential energy $V(Q)$ is barred from entering at all, to the order in
$t$ accepted. It follows that our result (\r{3.13}) for the coherence
norm is valid only in the limit when the decoherence times (\r{4.1}) are
much smaller than any of the time scales characteristic of the free
motions of the single-freedom system as well as the reservoir,
\be
\tau_{\rm dec}\ll \tau_{\rm sys},\,\tau_{\rm res}\,.
\l{4.2}
\ee 
Just for the sake of illustration, if the single-freedom system were an
oscillator the relevant system time scale would be the basic period of
oscillation, while for the reservoir the shortest time scale is either the
inverse of the highest frequency provided by the environmental degrees
of freedom or the thermal time $\hbar/kT$.
In Sect. \ref{shortmemory} we discuss the more general regime where no
assumption is made about the relative size of
$\tau_{\rm dec}$ and $\tau_{\rm res}$.

The selfconsistency condition (\ref{4.2}) is fulfilled for sufficiently
large distances between the superposed wave packets; it is hard, 
probably impossible, to violate for truly macroscopic superpositions and
that fact may be seen as the reason for the absence of quantum
interferences from the macroworld.
We hasten to add that the condition is not fulfilled for the present-day
experiments on decoherence which all still operate in the
situation $ \tau_{\rm res}\ll\tau_{\rm sys} \ll\tau_{\rm dec}$ where
the golden rule applies. In order to distinguish the decoherence
phenomena taking place in our short-time limit from the golden-rule type
decoherence processes thus far observed we speak of
``interaction dominated decoherence'' in the limit (\ref{4.2}).

Within its range of applicability, our short-time result has a certain
universal character. Inasmuch as the free-motion
Hamiltonian $H_{\rm sys}+H_{\rm res}$ is not operative through the full
cycle of any free oscillation in either the single-freedom system or the
reservoir, the character of such oscillations remains irrelevant for
decoherence as a short-time phenomenon. It does not matter whether the
single-freedom system is a harmonic or anharmonic oscillator since, as
already stressed before, the force $-V'(Q)$ gets no chance to act;
likewise, whether the bath consists of oscillators (like lattice
vibrations or the electromagnetic field), two-level atoms, or other
elementary units is immaterial.

The insensitivity of interaction-dominated decoherence to the character
of the oscillations generated by $H_{\rm sys}+H_{\rm res}$ also implies ignorance of whether the reservoir will, at larger times,
impose underdamped or overdamped motion to the single-freedom system.

\subsection{Wave Packets Distinguished by the Coupling Agent $Q$}

If the center positions $q_1,q_2$ of the superposed wave packets are
classically distinct the decay of the interference term
$\rho^{12}_{\rm sys}(t)$ is governed by the first of the three
exponentials, ${\cal E}^Q(t)=\exp\{-(t/\tau_{\rm dec}^Q)^2\}$, in the
norm $N_{12}(t)$. We then encounter a Gaussian fall-off on the time
scale $\tau_{\rm dec}^Q$. The second and third exponentials are
trivially ineffective for $p_1=p_2$; but even for $|p_1-p_2|\neq 0$ and
independent of $\hbar$ they may be considered as practically constant in
time since their life times $\tau_{\rm dec}^{QP}\propto
\hbar^{2/3},\tau_{\rm dec}^P\propto\hbar^{1/2}$ are much larger than
$\tau_{\rm dec}^Q\propto\hbar$. Equally ineffective are the corrections
of third and higher order in $t$ within that first exponential, in our
limit of large distances $|q_1-q_2|$. This is because the whole exponent
in ${\cal E}^Q(t)$ depends on the distance only through the common
factor $(q_1-q_2)^2$; the leading $t^2$-term thus defines a scaling
variable $\tau=t|q_1-q_2|$ such that higher-order corrections involve
$t^n=\tau^n/|q_1-q_2|^{n-2}$. For sufficiently large distances
$|q_1-q_2|$ the higher-order corrections would come into effect
only for times $t$ at which the leading Gaussian has already suppressed
the coherence norm to rather uninterestingly small values. For the same
reason the prefactor ${\cal P}(t)$, which arises from the Gaussian
integrals, cannot noticeably deviate from its initial value unity during
the life time of the leading exponential.

\subsection{Wave Packets Distinguished by the Conjugate Momentum $P$}

An interesting situation arises when $q_1=q_2$ and $|p_1-p_2|$ is of
classical magnitude, i.e. independent of Planck's constant, since then
the first and second exponentials in the coherence norm remain equal to
unity at all times. Our interaction dominated decoherence is thus
described by the third exponential, $N_{12}(t)={\cal
E}^P(t)=\e^{-(t/\tau_{\rm dec}^{P})^4}$. Due to the different power of
Planck's constant in $\tau_{\rm dec}^{P}$, i.e. $\tau_{\rm
dec}^{P}/\tau_{\rm dec}^{Q}\propto\hbar^{1/2}$, we may say that under
the influence of the position $Q$ as a coupling agent, momentum-space
superpositions decohere more slowly than position-space superpositions.
It has in fact been known for quite some time that an interaction
$H_{\rm int}=QB$ decoheres superpositions of wave packets most rapidly
if these packets are distinct in the eigenrepresentation of $Q$; Zurek
\cite{Zurek} speaks of the ``distinction of the pointer basis''; we here
see that distinction carrying over to our short-time limit of
decoherence which has previously received little attention, in spite of
its relevance for the emergence of classical behavior in the macroworld.
Part of the importance of our result (\ref{3.13}) lies in bringing to
light the rapid decay of superpositions of packets not at all
distinguished by the coupling agent $Q$. As already mentioned before,
the capability of $H_{\rm int}=QB$ to decohere momentum-space
superpositions would be overlooked if the action of the free-motion
Hamiltonian $H_{\rm sys}+H_{\rm res}$ were dropped entirely, with
overzealous appeal to the limit (\ref{4.2}) of interaction predominance. Clearly, our
short-time expansion of the (logarithm of the) interaction-picture
propagator $\tilde{U}(t)$ accounts, in the next-to-leading order in the
time $t$, for just that much free motion as necessary to let a
pure-momentum superposition acquire a bit of a $Q$ component and thus to
become visible and fall prey to $H_{\rm int}=QB$.

\subsection{Transition Between Position Space and Momentum Space
Superpositions}

The borderline between position-space and momentum-space distinction is
worth a moment of special attention. When both $|q_1-q_2|$ and
$|p_1-p_2|$ are nonzero and of classical magnitude (independent of
$\hbar$), the first exponential, ${\cal E}^Q(t)$,  with its Gaussian decay dominates the
decoherence process, as already emphasized above. Now imagine the
momentum distinction fixed and the distance $|q_1-q_2|$ decreased;
eventually, the life time $\tau_{\rm dec}^Q$ of the first exponential
will have grown to the magnitude of its competitors $\tau_{\rm
dec}^{QP},\;\tau_{\rm dec}^P$, and then the dominance of the first
exponential is lost. The emancipation of the competing exponentials
takes place when, respectively, $\tau_{\rm dec}^Q/\tau_{\rm
dec}^{QP}={\cal O}(\hbar^0)\approx 1$ and $\tau_{\rm dec}^Q/\tau_{\rm
dec}^P={\cal O}(\hbar^0)\approx 1$; inserting the various decoherence
times according to (\ref{4.1}) we see that both transitions concur at
\begin{equation}
  |q_1-q_2|^2/|p_1-p_2|={\cal O}(\hbar)\,,
  \label{4.3}
\end{equation}
i.e. for classical magnitude of the momentum distinction at
$|q_1-q_2|\propto \sqrt{\hbar}$. Interestingly, then, the transition
in question requires keeping all three exponentials in the coherence
norm (\ref{3.13}) for a proper description. Actually, to obtain good
quantitative reliability it would be advisable to include the
order-$t^2$ term in the prefactor ${\cal P}(t)$ as well,
${\cal P}(t)=\left(1+4\sigma \la B^2\ra t^2/\hbar^2+{\cal
O}(t^4)\right)^{-1/2}\approx\exp\{-2\sigma\la B^2\ra t^2/\hbar^2\}$,
since the position-space width of each of the superposed wave packets
was assumed as $\sqrt{\sigma}\propto\sqrt{\hbar}$, i.e. as of the same
order in $\hbar$ as the transitional distance $|q_1-q_2|$.

\section{Several Reservoirs and Coupling Agents}

A single-freedom system may be coupled to two many-freedom reservoirs
with both the position $Q$ and the momentum $P$ serving as system coupling
agents, according to the interaction Hamiltonian \cite{BHS}
\begin{equation}
  H_{\rm int}=QB_Q+PB_P\,.
  \label{5.1}
\end{equation}
The two separate reservoirs enter with the respective coupling agents
$B_Q,B_P$; for these we assume the structure (\ref{3.9}), i.e.
$B_Q=\sum_iB_{Qi},\,B_P=\sum_iB_{Pi}$, and vanishing means w.r.t. the
initial state of the reservoirs. The ``$Q$-reservoir'' and the
``$P$-reservoir'' are independent and have their own free-motion
Hamiltonians such that $H_{\rm res}=H_Q+H_P$.

To describe the decoherence of an initial superposition like
(\ref{2.2},\ref{2.3}) in the limit (\ref{4.2}) we may again employ the
short-time expansion of the (logarithm of the) interaction picture
propagator. In analogy to (\ref{3.4},\ref{3.7}) we have
\begin{equation}
  \tilde{U}(t)=\Big(\!\e^{-\i\int_0^t\!dt'\tilde{H}_{\rm
  int}(t')/\hbar}\Big)_{\!+} = \e^{-\i\{(QB_Q+PB_P)t+{\cal
  O}(t^2)\}/\hbar} \,.
  \label{5.2}
\end{equation}
Note that we need not go to higher than first order in $t$ since the
presence of both reservoirs entails the appearance of both the position {\it and}
the momentum in first order; no bit of free motion must be invoked here
to assist any underprivileged distinction of the superposed wave
packets.
The central limit theorem then yields Gaussian decay of the coherence norm,
\begin{eqnarray}
  N_{12}(t)&=&\exp\{-(t/\tau_{\rm dec}^Q)^2\}\,\exp\{-(t/\tau_{\rm dec}^P)^2\}
   \,,\nonumber  \\
  \tau_{\rm dec}^Q&=&\hbar\Big/|q_1-q_2|\sqrt{\la B_Q^2\ra}\,,\label{5.3}\\
  \tau_{\rm dec}^P&=& \hbar\Big/|p_1-p_2|\sqrt{\la B_P^2\ra} \,.
  \nonumber
\end{eqnarray}
The remarks about limits of validity and universality of the foregoing
section apply again, except for the simplification that the presence of both
reservoirs makes for symmetry between the pair of observables. In
particular, higher-order corrections in $t$ are irrelevant if at least
one of the two ``distances'' $|q_1-q_2|,\;|p_1-p_2|$ is of classical
magnitude.

\section{Competition of decoherence and bath correlation decay}
\label{shortmemory}

Thus far we have assumed that decoherence
is by far the fastest process, shorter even in duration than environmental
time scales such that the two conditions (\r{1.3}) and (\r{1.4}) could be
exploited. Of greater experimental relevance, however, is the case
in which (\r{1.3}) is satisfied, while the bath
correlation time scale $\tau_{\rm res}$ may be comparable with or even
shorter than the decoherence time. To that important case we shall now 
generalize our above discussions.

The analysis of Sect. \ref{IDD} goes through unchanged up to 
the short-time expansion (\ref{3.6}), except that this very expansion must now 
be confined to the free time evolution of the system coupling agent,
$\tilde{Q}(t)=Q+M^{-1}Pt+\ldots$, while
the time dependence of the bath coupling agent $\tilde{B}(t)$ generated by the
free bath Hamiltonian $H_{\rm res}$ must be kept in full,
\be
\tilde{H}_{\rm int}(t)=\tilde{Q}(t)\tilde{B}(t)=
\left(Q+\ldots\right)\tilde{B}(t)
\,.
  \label{7.1}
\ee
Actually, we shall simplify even further by confining ourselves to the 
lowest-order term of the expansion of the system coupling agent, $\tilde{Q}(t)
\approx Q$, thus confining ourselves to treating the decoherence of wave packets
with different locations in $Q$-space.

The propagator (\r{3.7}) is now replaced by
\begin{equation}
  \tilde{U}(t)=
\Big(\!\e^{-\frac{\i}{\hbar}\{Q\int_0^t ds \tilde{B}(s)+\ldots\}}\Big)_{\!+}\,.
  \label{7.2}
\end{equation}
Proceeding in the very same fashion as in Sect \r{IDD} we find the variant of 
(\r{3.8}),
\begin{eqnarray}
  \la q|\rho_{\rm sys}^{ij}(t)|q'\ra &=&D_Q(t)\,
\la q|\rho_{\rm sys}^{ij}(0)|q'\ra \l{7.3}\\D_Q(t)&=&
\Big\la\!
\Big(\!\e^{\frac{\i}{\hbar}\{q'\int_0^t ds \tilde{B}(s)\}}\Big)_{\!-}
\Big(\!\e^{-\frac{\i}{\hbar}\{q\int_0^t ds \tilde{B}(s)\}}\Big)_{\!+}
  \Big\ra \,.
\nonumber
\end{eqnarray}
Again, the large angular brackets 
denote an average w.r.t. the initial state of the reservoir, and
$(\ldots)_-$ refers to anti-time ordering, opposite in sense to $(\ldots)_+$.

As in Section \r{IDD} we now take advantage of the multi-component structure 
of the bath coupling agent $B$ which allows to regard
$\tilde{B}(t)$ as (an operator process) of Gaussian 
statistics. The reservoir average
in (\r{7.3}) may then be evaluated analytically (most straightforwardly by 
expanding all exponentials),
\ba
  & &\big\la\big(\e^{\frac{\i}{\hbar}q'\int_0^t ds \tilde{B}(s)}\big)_{\!-}
\big(\e^{-\frac{\i}{\hbar}q\int_0^t ds \tilde{B}(s)}\big)_{\!+}\,
  \big\ra=\l{7.4}\\& &\quad
\e^{-\frac{1}{\hbar^2}(q-q')\int_0^t ds \int_0^s ds'
\big(q\la \tilde{B}(s)\tilde{B}(s')\ra-q'\la
\tilde{B}(s')\tilde{B}(s)\ra\big)}\,.
  \nonumber
\ea
For the superposition of wave packets with
different positions studied here,
this result is a generalization of (\r{3.10}),
valid also for times
long compared to environmental correlation times.
A short time expansion of (\r{7.4}) recovers (\r{3.10}). We had previously 
\cite{BHS} derived (\r{7.4}) for the special case of a reservoir of harmonic 
oscillators; here, we can rejoice in the validity for general baths with 
effectively Gaussian coupling agents $B$.

All that remains to be done is to
determine the coherence norm $N_{12}(t)$ as
in Sect. \r{IDD}. That task, now simplified inasmuch as the momentum $P$
is barred, yields
\be
N_{12}(t) =
\exp\Big(\!-\frac{(q_1-q_2)^2}{\hbar^2}
\int_0^t \!\!ds \!\int_0^s \!\!ds' \,\la\{\tilde{B}(s),\tilde{B}(s')\}\ra\Big),
\label{7.5}
\ee
with no restriction on the validity beyond  $\tau_{\rm dec}\ll\tau_{\rm sys}$.
Clearly, the foregoing result generalizes the first factor 
${\cal E}^Q(t)$ in the coherence 
norm (\r{3.13}) so as to allow for competition of decoherence and bath 
correlation decay.

While we still observe the quadratic
dependence of the exponential suppression of coherence
on the distance $|q-q'|$, the precise time evolution
of decoherence is governed by the symmetric part of the
bath correlation function. No system time scale
is involved here, in contrast to the analogous expression
(\r{1.2}) for exponential Golden-Rule decay. In fact,
(\r{7.5}) describes non-exponential decay for $t\rightarrow \infty$,
unless the Fourier transform of
$\la\{\tilde{B}(t),\tilde{B}(0)\}\ra$ (the
spectral density), differs from zero at zero frequency.
This is seen by writing
$\int_0^t ds \int_0^s ds' \la\{\tilde{B}(s),\tilde{B}(s')\}\ra
=\int_0^t ds (t-s) \la\{\tilde{B}(s),\tilde{B}(0)\}\ra$,
taking advantage of the stationarity of the Gaussian process.
As $t\rightarrow \infty$, no rate of decay can be
defined, unless
$\int_0^\infty ds \la\{\tilde{B}(s),\tilde{B}(0)\}\ra$ remains
finite.
Examples of such decay will be presented for the exactly
solvable harmonic oscillator model in the accompanying paper
\cite{oscillator}.

\section{Angular-Momentum Decoherence}

To emphasize 
the universality of interaction dominated decoherence  
$\tau_{\rm dec}\ll \tau_{\rm sys},\,\tau_{\rm res}$ we here 
consider an angular momentum vector $\vec{J}$ whose three 
components obey the commutation relations $[J_x,J_y]=\i\hbar J_z$ 
etc, coupled to a reservoir. As the Hamiltonian we take
\be
H_{\rm sys}=\Omega J_z,\quad H_{\rm int}=J_xB\,.
\l{6.1}
\ee
The squared angular momentum is thus conserved, $\vec{J}^2=j(j+1)$, 
with the quantum number $j$ capable of taking on integer or half 
integer values; large values of $j$ enable the angular momentum to 
near classical behavior.

Suitable wave packets are provided by coherent states 
\cite{Angularcoherent} which specify a direction for (the expectation 
value of) $\vec{J}$ in terms of two angles, $\theta$ and $\phi$, 
with the minimal uncertainty allowed by the commutation relations. 
We shall denote those states by $|j,\theta,\phi\ra\equiv |\alpha\ra$, 
the latter shorthand dropping the quantum number $j$ and introducing 
the complex amplitude $\alpha=\e^{\i\phi}\tan(\theta/2)$. The whole
complex plane is visited by $\alpha$ as the ``polar'' angle ranges 
in $0\leq \theta\leq \pi$ and the ``azimuthal'' angle in $0\leq\phi<2\pi$. 
(We may speak of the mapping of the surface of the unit sphere onto 
the complex plane; the sphere $\lim_{j\to \infty}\vec{J}^2/(\hbar j)^2=1$ is 
the classical phase space.) The coherent-state mean of $\vec{J}$ reads
\ba
\la \alpha|J_x|\alpha\ra &=&\hbar j\,\f{\alpha+\alpha^*}{1+\alpha\alpha^*}
=\hbar j\cos\phi\sin\theta\,,\nonumber\\
\la\alpha|J_y|\alpha\ra &=&\hbar j\,\f{\i(\alpha^*-\alpha)}{1+\alpha\alpha^*}
=\hbar j\sin\phi\sin\theta
\l{means}\\
\la\alpha|J_z|\alpha\ra &=&\hbar j\,\f{1-\alpha\alpha^*}{1+\alpha\alpha^*}
=\hbar j\cos\theta
\nonumber\,.
\ea

The coherent state $|\alpha\ra$ can be expressed in terms of eigenstates 
$|j,m\ra$ of $\vec{J}^2$ and $J_z$ (eigenvalues $j(j+1)$ and $m$, 
respectively) as
\ba
|\alpha\ra&=&(1+\alpha\alpha^*)^{-j}\,\e^{\alpha J_-/\hbar}|j,j\ra
\nonumber\\
          &=&(1+\alpha\alpha^*)^{-j}\sum_{n=0}^{2j}\sqrt{2j\choose n}
\alpha^n|j,j-n\ra
\l{cohst}\\
          &\equiv& (1+\alpha\alpha^*)^{-j}||\alpha\ra\nonumber\,,
\ea
where $J_-=J_x-\i J_y$. It will in fact be convenient to work with 
the non-normalized Dirac ket $||\alpha\ra$ which is holomorphic in 
$\alpha$ and the corresponding antiholomorphic bra $\la\alpha||$. 
The coherent state $|\alpha\ra$ itself is normalized as 
$\la\alpha|\alpha\ra=1$.

We now turn to a superposition of two coherent states, 
$|\ra=c_\alpha|\alpha\ra+c_\beta|\beta\ra$, which in the limit 
$j\gg 1$ is a macroscopic superposition, and inquire about the 
temporal fate of the coherence norm 
$N_{\alpha\beta}(t)={\rm Tr_{sys}}\rho_{\rm sys}^{\alpha\beta}(t)
\rho_{\rm sys}^{\beta\alpha}(t)$ with
\ba
\rho_{\rm sys}^{\alpha\beta}(t)&=&
\Big((1+\alpha\alpha^*)(1+\beta\beta^*)\Big)^{-j}\nonumber\\
& &\times\,{\rm Tr_{res}}\,
\e^{-\i Ht/\hbar}
||\alpha\ra\la\beta||\,\rho_{{\rm res}}(0)\,\e^{\i Ht/\hbar}
\l{coh1}
\ea 
the temporal successor of $\rho_{\alpha\beta}(0)=|\alpha\ra\la\beta|$.
Like in Sect.\r{IDD} we go to the interaction picture where
\ba
\tilde{H}_{\rm int}&=&\tilde{B}(t)\left(J_x\cos\Omega t-J_y\sin\Omega t\right)
\label{Hint}\\
&=&J_xB+(J_x\dot{B}-\Omega J_yB)t\nonumber\\
& &+(-\Omega^2J_xB-2\Omega
J_y\dot{B}+J_x\ddot{B})t^2/2+\ldots\nonumber
\ea
gives rise to the propagator
\ba
\tilde{U}(t)&=&\exp\left\{-\textstyle{\f{\i}{\hbar}}
(BJ_xt-(\dot{B}J_x-B\Omega J_y)t^2/2\right. \label{short}\\
& &\quad\quad +\left(-\Omega^2J_xB-2\Omega J_y\dot{B}+J_x\ddot{B}
\right.\nonumber\\
& &\qquad\quad\;\left.\left.
+\textstyle{\frac{\i}{2\hbar}}[B,\dot{B}]+\textstyle{\frac{1}{2}}\Omega
J_zB^2\right)t^3/6
+\ldots\right\}\,; \nonumber
\ea
see the appendix for the derivation of the foregoing short-time
expansion. Note that we have here included the third-order term of the
expansion, for a reason that will become clear presently.
When the propagator $\tilde{U}(t)$ acts on the holomorphic $||\alpha\ra$
we may use the identities
\ba
J_x||\alpha\ra&=&\f{\hbar}{2}\left(2j\alpha-(\alpha^2-1)\f{\partial}
{\partial\alpha}\right)
||\alpha\ra\equiv \hat{X}_\alpha||\alpha\ra\nonumber\\
J_y||\alpha\ra&=&\f{\hbar}{2\i}\left(2j\alpha-(\alpha^2+1)\f{\partial}
{\partial\alpha}\right)
||\alpha\ra\equiv \hat{Y}_\alpha||\alpha\ra \label{holids} \\
J_z||\alpha\ra&=&\hbar\left(j-\alpha\f{\partial}{\partial\alpha}\right)
||\alpha\ra\equiv \hat{Z}_\alpha||\alpha\ra
\nonumber
\ea
and their adjoints
$\la\beta||J_x=X_{\beta^*}\la\alpha||$ etc. such that
$U(t)||\alpha\ra=U(t;\alpha)||\alpha\ra$ with $U(t;\alpha)$ differing
from $U(t)$ only by the replacements (\ref{holids}); similarly,
$\la\beta||U^\dagger(t)=U^\dagger(t,\beta^*)\la\beta||$ with
$U^\dagger(t,\beta^*)$ obtained from $U^\dagger(t)$ by $J_x\to
X_{\beta^*}$ etc..  We thus get
\ba
& &\tilde{\rho}_{\rm sys}^{\alpha\beta}(t)
\Big((1+\alpha\alpha^*)(1+\beta\beta^*)\Big)^{j}\nonumber\\& &\hspace{.5cm}=
{\rm Tr_{res}}\,U(t)
||\alpha\ra\la\beta||\,\rho_{{\rm res}}(0)\,U^\dagger(t)\nonumber\\
& &\hspace{.5cm}=
{\rm Tr_{res}}\,U^\dagger(t,\beta^*)U(t,\alpha
||\alpha\ra\la\beta||\,\rho_{{\rm res}}(0)\label{coh}\\
& &\hspace{.5cm}=
\la U^\dagger(t,\beta^*)U(t,\alpha)\ra\,
||\alpha\ra\la\beta||\nonumber\,.
\ea
To within a further correction of order $t^4$ we can merge the two 
exponentials in the last member of the foregoing equation  by simply 
adding the exponents. We proceed to the coherence norm
\ba
& &N_{\alpha\beta}(t)=\Big((1+\alpha\alpha^*)(1+\beta\beta^*)\Big)^{-2j}\,
\nonumber\\
& &\times\left\la\exp\left(-\i\left\{(\hat{X}_\alpha-\hat{X}_{\beta^*})
(Bt+\dot{B}t^2/2+(\ddot{B}-\Omega^2B)t^3/6)\right.\right.\right.\nonumber\\
& &\hspace{5em} -(\hat{Y}_\alpha-\hat{Y}_{\beta^*})
(Bt^2/2+\dot{B}t^3/3)\nonumber\\
& &\hspace{5em} +(\hat{Z}_\alpha-\hat{Z}_{\beta^*})B^2t^3/12\label{cohnorm}\\
& &\hspace{5em}
\left.\left.\left.+(\hat{X}_\alpha^2-\hat{X}_{\beta^*}^2)
\textstyle{\f{\i}{\hbar}}
[B,\dot{B}]t^3/12\right\}\right)\right\ra\nonumber\\
& &\times\Big\la\exp\Big(-\i\Big\{\,{\rm same\; with\;}\alpha\to\beta,\,
\beta^*\to\alpha^*\Big\}\Big)\Big\ra\nonumber\\
& &\times \Big((1+\alpha\alpha^*)(1+\beta\beta^*)\Big)^{2j}\nonumber\,,
\ea
where we have encountered ${\rm Tr_{sys}}||\alpha\ra\la\beta||\beta\ra
\la\alpha||=\Big((1+\alpha\alpha^*)(1+\beta\beta^*)\Big)^{2j}$; it is
on this latter function of $\alpha,\alpha^*,\beta,\beta^*$ that the 
various differential operators like
$\partial/\partial \alpha$ in the exponentials in (\r{cohnorm}) act. To
leading order in $j$ these differentiations act as
$\partial/\partial\alpha\to 2j\alpha^*/(1+\alpha\alpha^*)$ etc., whereupon the
differential operators $\hat{X}_\alpha,\hat{Y}_{\alpha},\hat{Z}_{\alpha}$ 
become replaced by real
c-numbers, in fact the coherent-state expectation values of $J_x,J_y,J_z$
given in (\r{means}), $\hat{X}_\alpha\to\la\alpha|J_x|\alpha\ra/\hbar$,
$\hat{X}_{\beta^*}\to\la\beta|J_x|\beta\ra/\hbar$ etc.. The two reservoir 
means  $\la\exp(\ldots)\ra$ in the foregoing expression for the coherence 
norm thus become mutual complex conjugates and are controlled by the three 
``distances''
\be
d_i=\la\alpha|J_i|\alpha\ra-\la\beta|J_i|\beta\ra\,,\quad 
i=x,y,z\,,
\l{6.10}
\ee 
and even by the differences of the mean values of $J_x^2$ wrt to 
the coherent states $|\alpha\ra,\,|\beta\ra$. Confining ourselves to the 
leading order in $j$ we have 
\ba
N_{\alpha\beta}(t)&=&
\Big|\!\Big\la\!\exp\!\Big(\!-\textstyle{\f{\i}{\hbar}}\Big\{d_x
\big(Bt+\dot{B}t^2/2+(\ddot{B}-\Omega^2B)t^3/6\big)
\nonumber\\
& &\hspace{5em} -d_y
\Big(Bt^2/2+\dot{B}t^3/3\Big)\nonumber\\
& &\hspace{5em} +d_z
B^2t^3/12\label{cohnormfin}\\
& &
\left.\left.\left.+\big(\la\alpha|J_x|\alpha\ra^2-\la\beta|J_x|\beta\ra^2\big)
\textstyle{\f{\i}{\hbar}}
[B,\dot{B}]t^3/12\right\}\right)\right\ra\!
\Big|^2\nonumber \,,
\ea
and this can now be seen to imply a greater wealth of decoherence courses 
than previously encountered for a canonical pair of observables.

The system coupling agent, $J_x$ in the interaction (\r{6.1}), again plays a 
distinguished role; it is most efficient in decohering a superposition 
$c_\alpha|\alpha\ra+c_\beta|\beta\ra$ if it has macroscopically distinct 
means in the two 
superposed coherent states, macroscopic now meaning $j\gg1$. In that situation
only the single term linear in the time $t$ need to be kept in the exponent 
of the coherence norm (\r{cohnormfin}). The Gaussian average for the bath 
coupling agent $B$ then yields 
$N_{\alpha\beta}(t)=|\la\e^{-\i d_xBt/\hbar}\ra|^2=
\e^{-(t/\tau_{\rm dec}^x)^2}$ with the decoherence time
\ba
\tau_{\rm dec}^x &=&d_x\sqrt{\la B^2\ra}/\hbar\l{taux}\\
&=&\left(\la B^2\ra \,j^2
\left( \cos\phi_\alpha\,\sin\theta_\alpha
-\cos\phi_\beta\,\sin\theta_\beta \right)^2\right)^{-\f{1}{2}}
\nonumber
\ea
in analogy with $\tau_{\rm dec}^Q$ of \r{4.1}. 

The competing terms in the coherence norm can become 
effective only when the distance $d_x$ vanishes (or is of 
subclassical magnitude). This happens in four distinct cases, three of which 
come with $\cos\phi_\alpha=\cos\phi_\beta,\,
\sin\theta_\alpha=\sin\theta_\beta$: (i) $\beta=1/\alpha^*\Longleftrightarrow
\{\phi_\alpha=\phi_\beta,\,
\theta_\alpha=\pi-\theta_\beta\}$ such that the two points in the spherical 
phase space distinguished by  $\alpha$ and $\beta$ are reflections of 
one another in the equatorial plane $\theta=\pi/2$; (ii) $\beta=\alpha^*
\Longleftrightarrow\{\phi_\alpha=
2\pi-\phi_\beta,\,\theta_\alpha=\theta_\beta\}$ whereupon the two points 
are mutually opposite on the circular section of the spherical phase space 
with the plane $\theta=\theta_\alpha=\theta_\beta$; (iii) $\beta=1/\alpha
\Longleftrightarrow\{\phi_\alpha=
2\pi-\phi_\beta,\,\theta_\alpha=\pi-\theta_\beta\}$ and then the two 
points are mutual antipodes. A fourth case, (iv), arises from $\cos\phi_\alpha=
\sin\theta_\beta,\,\cos\phi_\beta=\sin\phi_\alpha$.  
At any rate, if $d_x=0$ but $d_y$ is of classical magnitude we may drop all 
terms of order $t^3$ in the coherence norm and get 
$N_{\alpha\beta}(t)=|\la\e^{-\i d_yB\Omega t^2/2\hbar}\ra|^2=
\e^{-(t/\tau_{\rm dec}^y)^4}$ with a decoherence time much larger than 
$\tau_{\rm dec}^x$,
\ba
\tau_{\rm dec}^y&=&\left(d_y^2\,\Omega^2\la B^2\ra/4\hbar^2\right)^{-\f{1}{4}}
\l{tauy}\\&=&
\left(\textstyle{\f{1}{4}}\,j^2\Omega^2\la B^2\ra 
\left( \sin\phi_\alpha\,\sin\theta_\alpha
-\sin\phi_\beta\,\sin\theta_\beta \right)^2\right)^{-\f{1}{4}}\,,
\nonumber
\ea
in analogy with $\tau_{\rm dec}^P$ of \r{4.1}.  Such  ``protection 
of coherence by symmetry'' has been discussed previously in 
Ref. \cite{BraunHaa}, in the context of golden-rule type decoherence

Specific to the angular-momentum algebra is the possibility that both $d_x$ 
and $d_y$ vanish but $d_z\neq0$; this actually happens in case (i) above as 
well as in the subcase $\cos(\phi_\alpha\pm\theta_\alpha)=0$ of case (iv). 
We then get the coherence norm, after doing 
a slightly different Gaussian integral, as 
$N_{\alpha\beta}(t)=|\la \e^{\{-\i d_zB^2t^3/12\hbar\}}\ra|^2
=\left(1+(t/\tau_{\rm dec}^z)^6\right)^{-\f{1}{2}}$\,;
the pertinent time scale is
\ba
\tau_{\rm dec}^z&=&\left(d_z^2\,\Omega^2\la B^2\ra^2/36\hbar^2\right)
^{-\f{1}{6}}\l{tauz}\\&=&
\left(\textstyle{\f{1}{36}}\,j^2\Omega^2\la B^2\ra^2 
\left(\cos\theta_\alpha
-\cos\theta_\beta \right)^2\right)^{-\f{1}{6}}
\nonumber
\ea

We refrain from a detailed discussion of the various transitional regimes 
that may arise when $d_x$ and $d_y$ are not strictly zero but of subclassical 
magnitude, a discussion that would proceed much in analogy to the one in 
Sect. IV. 

We would like to emphasize that the decoherence times 
$\tau_{\rm dec}^{x,y,z}$ all obey the power law (\r{1.1}), with $1/j$ as a 
dimensionless representative of Planck's constant; the exponents tend 
to order the decoherence times in magnitude as $\tau_{\rm dec}^x\ll
\tau_{\rm dec}^y\ll\tau_{\rm dec}^z$; that ordering expresses decreasing power
of the coupling agent $J_x$ in decohering the respective superpositions. As 
usual, the coupling agent is most effective with respect to superpositions of 
states it ``sees'' as distinct in terms of its respective mean values; next 
come superpositions of states distinct by the mean values of $J_y$ since the 
free motion generated by $H_{\rm sys}=\Omega J_z$ rotates $J_x$ into $J_y$; 
finally, superpositions of states distinguished only by $J_z$ undergo slowest
decoherence since $J_z$ enters the short-time expansion of the propagator
$(\exp\{-i\int_0^tdt'\tilde{H_{\rm int}(t')/\hbar}\})_+$ only in the 
third-order term $t^3$, due to the commutator $[J_x,J_y]=\i\hbar J_z$. 

The partial immunity to decoherence of superpositions of 
angular-momentum coherent states expressed in the ordering just discussed
may be broken by reducing the symmetry 
of the dynamics. One way of achieving that is to generalize the free 
motion as 
$H_{\rm sys}=\Omega_zJ_z+\Omega_yJ_y$; another is to allow for 
more reservoirs \cite{BHS}, e.g. according to $H_{\rm int}=J_xB_x+J_zB_z$.

Clearly, a larger set of observables like the generators of, say, SU($n$) with
$n=3,4,\ldots$ would give rise to a yet richer decoherence scenario if 
$H_{\rm sys}$ and $H_{\rm int}$ both linearly involved different such 
generators.

\section{Conclusions and Perspectives}

Quantum superpositions are fragile objects with respect to 
almost all environmental influences. 
In quantum mechanics, ``openness'' of a system is a more involved concept
than in classical mechanics.
Although good isolation from the environment
may allow damping to
hardly be noticeable for quantities with a classical limit,
coherences in a quantum system may
be subjected to rapid decay. The underlying time scale separation
between $\tau_{\rm dec}$ and $\tau_{\rm diss}$
becomes ever more drastic as the ``distance'' between the superposed 
states grows. The decoherence time scale is shortened by
a factor involving the distance, measured in
units of a quantum reference ``length''
and thus enormously big when it comes to mesoscopic
or even macroscopic scales.
For more and more
macroscopic superpositions, the decoherence time scale
eventually becomes the smallest time scale involved.
It follows that 
standard approaches to open system dynamics, based on
Golden-Rule-type assumptions fail to describe the
rapid decay of such superpositions.

We have shown that a short-time expansion of
the logarithm of the interaction propagator is the appropriate 
approach to decoherence in the limit of macroscopic superpositions.
Remarkably, decoherence dynamics in this limit is 
largely independent of the nature of the system and the bath.
No classical forces will have time to exert their influence
on the very short decoherence time scale.

A remark about the use of a factorized initial condition is
in order: Our results ignore the problem of how to actually
create macroscopic superpositions. We assume they
are given and determine the ensuing dynamics.
Clearly, under laboratory conditions, it will take
a certain time to prepare such an initial state,
time enough for decoherence to possibly be effective.
Initial system-environment correlations are thus an
important ingredient for the discussion of the decay of
macroscopic superpositions, a problem that will be
addressed in future work. 

How far the creation of superpositions can be stretched
to the macroscopic is a question of central importance not only for quantum 
foundations but also for engineering
in the fields of quantum information.
Our results suggest that for these fascinating
developments  environmental effects 
need to be described with new theoretical input. Well established 
methods of open-system dynamics, 
historically developed 
with an eye to near-equilibrium behavior
become questionable for the 
non-equilibrium dynamics of coherent
phenomena and may well
turn out to be too limited to meet the 
quantum challenges of the future.

\section*{Acknowledgments}

We have enjoyed discussions with Wojciech Zurek, Dieter Forster, 
and Hans Mooij as well as
the hospitality of the Institute for Theoretical Physics at the
University of Santa Barbara during the workshop
'Quantum Information: Entanglement, Decoherence, Chaos', 
where this project was completed.
Support by the Sonderforschungsbereiche
``Unordnung und Gro{\ss}e Fluktuationen'' and
``Korrelierte Dynamik hochangeregter atomarer und molekularer Systeme''
of the Deutsche Forschungsgemeinschaft is gratefully acknowledged.

\appendix\label{appendix}

\section{Short-Time Expansion}

To derive the expansions (\ref{3.7},\ref{short}) of the interaction-picture
propagator  we start with expanding the interaction Hamiltonian
(\ref{3.6}),
\begin{eqnarray}
  \tilde{H}_{\rm int}(t)=H_0+H_1t+H_2t^2/2+{\cal O}(t^3)\,. \label{A.1}
\end{eqnarray}
Separating the time independent term $H_0$ we write the propagator
\begin{eqnarray}
\tilde{U}(t)&=& \left(\e^{-\i\int_0^tdt'\tilde{H}_{\rm
int}(t')/\hbar}\right)= \e^{-\i H_0t/\hbar} U_1(t)\nonumber \\
U_1(t)&=&\left(\e^{-\i\int_0^tdt'H_1(t')/\hbar}\right)_+
\label{A.2}\\
H_1(t)&=&\e^{\i H_0t/\hbar}\left(H_1t+H_2t^2/2+{\cal O}(t^3)\right)
\e^{-\i H_0t/\hbar}\nonumber\\
&=&H_1t+\frac{\i}{\hbar}[H_0,H_1]t^2+H_2t^2/2+{\cal O}(t^3)\nonumber \,.
\end{eqnarray}
Next, we split off the leading term $H_1t$ in $H_1(t)$,
\begin{eqnarray}
 U_1(t)&=&\e^{-\i H_1t^2/2\hbar}U_2(t) \nonumber\\
 U_2(t)&=&\left(\e^{-\i\int_0^tdt'H_2(t')/\hbar}\right)_+
\label{A.3}\\
H_2(t)&=&\e^{\i
H_2t^2/2\hbar}\left((\textstyle{\frac{2\i}{\hbar}}[H_0,H_1]+H_2)t^2/2)\right)
\e^{-\i H_1t^2/2\hbar}\nonumber\\
&=&\left(\textstyle{\frac{2\i}{\hbar}}[H_0,H_1]+H_2\right)t^2/2+{\cal O}(t^3)
\,,\nonumber \\
\Longrightarrow U_2(t)&=&\e^{-\i
\left(2\i[H_0,H_1]/\hbar+H_2\right)t^3/6+{\cal O}(t^4)}\nonumber\,.
\end{eqnarray}
When finally merging the three unitary factors
$\e^{-\i H_0t/\hbar}U_1(t)U_2(t)$
into a single exponential we encounter a correction of the $t^3$ term
due to
\begin{equation}
  \e^{-\i H_0t/\hbar}\e^{-\i H_1t^2/2\hbar}=
  \e^{\{-\f{\i}{2\hbar}(H_0t+H_1t^2-\f{\i}{4\hbar}[H_0,H_1]t^3+{\cal
  O}(t^4))\}}\,,
  \label{A.4}
\end{equation}
whereupon we get
\begin{equation}
  \tilde{U}(t) =\e^{\{-\f{\i}{\hbar}(H_0t+H_1t^2/2+(2H_2+\f{\i}{\hbar}
[H_0,H_1])t^3/12+{\cal
  O}(t^4))\}} \,.
  \label{A.5}
\end{equation}

The foregoing general identity yields (\ref{3.7}) since the interaction
Hamiltonian (\ref{3.6}) implies $H_0=QB$ and $H_1=M^{-1}PB+Q\dot{B}$.
For the angular momentum case, where we needed the third-order term in
$\ln \tilde{U}(t)$ to reveal the quantum acceleration of decoherence for
the most obstinate superpositions of coherent states.


\begin{references}
\bibitem{Fuss} Our power counting assumes $\la B^2\ra$
independent of $\hbar$; if $\la B^2\ra\propto \hbar^\gamma$,
with $\gamma<2$,
as e.g. in the zero temperature harmonic oscillator case,
we still retain $\tau_{\rm dec}\to 0$ for $\hbar\to 0$, see
\cite{oscillator}.
\end{references}
\end{document}